\documentclass[12pt]{article}
\usepackage{%
pstricks,%
pst-plot,%
epsf%
}                                  
\usepackage{cite}
\usepackage{array,dcolumn}	
\usepackage{axodraw}
\newcommand{\be}{\begin{equation}}
\newcommand{\ee}{\end{equation}}
\newcommand{\bea}{\begin{eqnarray}}
\newcommand{\eea}{\end{eqnarray}}
\newcommand{\baa}{\begin{eqnarray*}}
\newcommand{\eaa}{\end{eqnarray*}}
\newcommand{\bet}{\begin{center} \begin{tabular}}
\newcommand{\ent}{\end{tabular} \end{center}}
\newcommand{\bary}{\begin{array}}
\newcommand{\eary}{\end{array}}
\newcommand{\bit}{\begin{itemize}}
\newcommand{\eit}{\end{itemize}}
\newcommand{\ra}{\rightarrow}

\newcommand{\crn}{\nonumber \\}
\newcommand{\non}{\nonumber}
\newcommand{\nn}{\nonumber}
\newcommand{\noi}{\noindent}

\newcommand{\STis}{Slavnov-Taylor identities }

\newcommand{\gv}{\mbox{GeV}}

\newcommand{\MS}{${\mathrm{MS}}$ }
\newcommand{\MOM}{${\mathrm{MOM}}$ }

\newcommand{\BFMOM}{${\mathrm{BF-MOM}}$ }
\newcommand{\BFMOMm}{{\mathrm{BF-MOM}} }
\newcommand{\MSb}{$\overline{\mathrm{MS}}$ }
\newcommand{\MSbm}{\overline{\mathrm{MS}} }

\newcommand{\Li}[1]{\,{\rm Li}_{#1}}

\newcommand{\bb}{\begin{thebibliography}}
\newcommand{\eb}{\end{thebibliography}}
\newcommand{\ci}[1]{\cite{#1}}
\newcommand{\bi}[1]{\bibitem{#1}}

\newfont{\liste}{pzdr scaled 1100}

\begin{document}
\thispagestyle{empty}
\onecolumn
\date{\today}
\vspace{-1.4cm}
\begin{flushleft}
{DESY 98-093 \\}
{hep-ph/9809485\\}
September, 1998

\end{flushleft}
\vspace{1.5cm}

\begin{center}

{\LARGE {\bf
  Exact mass dependent two--loop $\overline{\alpha}_s(Q^2)$  
 in the
  background \MOM renormalization scheme
        }
}
\vspace{1.5cm}

\vfill
{\large
 F.~Jegerlehner~~and~~O.V.~Tarasov\footnote{
 {~On leave of absence from JINR, 141980 Dubna (Moscow Region),
 Russian
 Federation.}}
}

\vspace{2cm}
Deutsches Elektronen-Synchrotron DESY\\
Platanenallee 6, D--15738 Zeuthen, Germany\\

\vfill
\end{center}
\begin{abstract}
A two-loop calculation of the renormalization group 
$\beta$--function in a momentum subtraction scheme with 
massive quarks is presented
using the background field formalism. The results have been obtained
by using a set of new generalized recurrence relations proposed
recently by one of the authors (O.V.T.).  The behavior of the mass
dependent effective coupling constant is investigated in detail.
Compact analytic results are presented.
\end{abstract}

\vfill
\newpage

\setcounter{footnote}{0}

\section{Introduction}
Particle masses are amongst the most important physical parameters and
in many cases their meaning and definition by thresholds (e.g. lepton
masses), symmetry breaking parameters (current quark masses, neutrino
masses) or scale parameters is quite clear. For particles which exist
as free or quasi--free states a definition by the pole mass is most
natural and has an unambiguous meaning. The definition of quark
masses, in particular for the light quarks, allows for a lot of
freedom, mainly because the pole mass is not directly observable due
to the confinement property of QCD. Nevertheless, quark masses play a
crucial role for the effective behavior of strong interactions at a
given scale. The purpose of the present calculation is a precise
understanding of the quark mass dependence of QCD, more specifically,
of the effective coupling constant $\alpha_s(Q^2)=g_s^2/(4\pi)$, the
most important quantity in the description of strong interactions.
These considerations are important for a better understanding of the
decoupling of heavy particles and of the relationship between QCD with
massive quarks and QCD in the \MSb scheme where effective theories
with different number of (light)
flavors~\cite{DecouplingWX,DecouplingBW,DecouplingM} must be matched
at the different quark thresholds.

When the on-shell renormalization scheme is not adequate we either may
use a minimal subtraction scheme (\MS or \MSb) or some version of a
momentum subtraction scheme (\MOM ), defined by the condition that
the radiative corrections of an appropriate set of quantities vanish
at a certain (off--shell) momentum configuration. While the
\MSb~\cite{MSb} scheme is technically simple and respects the \STis
the \MOM scheme is more physical since it respects the decoupling
theorem~\cite{AC}. A serious short coming of the standard \MOM
schemes~\cite{GP}, however, is the fact that they spoil the validity
of the canonical form of the Slavnov-Taylor identities. An elegant way
out of this difficulty is the use of the so called background field
method (BFM)~\cite{Witt}. The latter takes advantage of the freedom to
chose a gauge fixing function in a particular way, namely, such that
the canonical \STis remain valid also after momentum subtractions. The
gauge invariant physical quantities are not affected by the gauge
fixing, however, the ``background field gauge'' selects a particular
representative of the gauge variant off--shell amplitudes. The
restauration of the \STis in the BFM is achieved solely by changing
the vertices with external gluons appropriately. For further details
and for the Feynman rules of QCD in the background field (BF)
formalism we refer to~\cite{Abbott,Rebhan}.

In Ref.~\cite{Rebhan} the renormalization group (RG) $\beta$--function of
QCD was evaluated at one--loop order in the background field approach
using the \MOM scheme.  In the present article we extend this analysis
to a complete mass dependent two--loop calculation. Previously, the
two-loop renormalization of the pure Yang-Mills theory in the
background field method was first considered in~\cite{Abbott} using
the \MSb scheme. Fermionic contributions were added
later in~\cite{AGS}. Calculations in the background formalism for an
arbitrary value of the gauge parameter were presented in~\cite{CM}. In
the standard approach the evaluation of the mass dependent QCD
$\beta$--function at the two-loop level was performed in~\cite{YH}. In
the latter publication only an approximate expression for the 
two--loop coefficient was given. Because of the complexity of such
calculations the result of~\cite{YH} was not confirmed by any other
group until now. The background field method provides the easiest way
to calculate a mass dependent $\beta$--function because here only
propagator diagrams need to be evaluated. A general method for the
evaluation of two-loop propagator type diagrams with arbitrary masses
was recently proposed in~\cite{OVT97,connection}.

Our paper is organized as follows: in Sec.~2 we describe the
calculation of the background field propagator, from which we obtain
the RG and the effective running coupling in the \BFMOM scheme in
Sec.~3. The relationship between the \BFMOM and the \MSb coupling is
presented in Sec.~4 (analytical) and Sec.~5 (numerical). For
completeness we include a formula for the bare BF propagator in
Appendix A. The BF Feynman rules and the BF propagator diagrams are
included in Appendices B and C, respectively.

\section{The background field propagator}
To regularize divergences we will use the dimensional regularization
procedure in $d=4-2\varepsilon$ dimensions. In the background field
approach, we only need to calculate the background field
renormalization constant $Z_A$ in order to obtain the charge
renormalization constant. The complete list of two-loop diagrams as
well as the Feynman rules in the background field approach may be
found in~\cite{Abbott} -- \cite{CM}.  $Z_A$ is determined by renormalizing
the background field propagator according to
\begin{equation}
\frac{1}{1+\Pi(Q^2,\mu^2,\{m_i^2\})}=\frac{Z_A}{1+\Pi_0(Q^2,\mu^2,\{m_i^2\})}
\;,
\label{za}
\end{equation}
where $Q^2=-q^2$ and $\mu$ is the subtraction point. Bare quantities
carry an subscript $_0$. In the \MOM scheme
the condition
\begin{equation}
\Pi(Q^2,\mu^2,\{m_i^2\})\left|_{Q^2=\mu^2}=0 \right.
\end{equation}
is imposed on the renormalized self--energy function.  The
renormalized mass $m_i$ in our calculations is defined as a pole of
the quark propagator. In the \MOM scheme $Z_A$ and therefore also the
RG $\beta$--function depend on the gauge parameter
$\xi$. The gauge parameter is renormalized by
\begin{equation}
\xi_0=\xi Z_3\;,
\end{equation}
where $Z_3$ is the renormalization constant of the quantum
gluon field. To circumvent problems connected with the renormalization
of the gauge parameter we have chosen the Landau gauge $\xi=0$.

We repeated the calculation of all two-loop diagrams in the background
formalism keeping non--vanishing quark masses. All calculations have
been performed with the help of FORM~\cite{FORM} using the algorithm
described in~\cite{OVT97}. Our results agree with those presented
in~\cite{Abbott,AGS,CM} for the limit of massless quarks. The sum of
all unrenormalized diagrams for an arbitrary value of the gauge
parameter is given in the Appendix.

The renormalized self--energy amplitude $\Pi(Q^2)$ has the form:
\begin{equation}
\Pi(Q^2)=\left( \frac{\alpha_s}{4\pi} \right) U_1
+\left( \frac{\alpha_s}{4\pi} \right)^2 U_2 + \cdots
\end{equation}
where
\begin{eqnarray}
&& U_1(Q^2/\mu^2,\{m_i^2/\mu^2\}) =\frac{11}{3}C_A \ln  \frac{Q^2}{\mu^2}
 +T_F \sum_{i=1}^{n_F}\left(
  \Pi_1\left(\frac{Q^2}{m_i^2}  \right)
 -\Pi_1\left(\frac{\mu^2}{m_i^2}\right)
                    \right),
  \nonumber \\
&& U_2(Q^2/\mu^2,\{m_i^2/\mu^2\})=\frac{34}{3}C^2_A \ln \frac{Q^2}{\mu^2}
 +T_F \sum_{i=1}^{n_F}\left( \Pi_2\left(\frac{Q^2}{m_i^2}  \right)
                          -\Pi_2\left(\frac{\mu^2}{m_i^2}\right)
		    \right).
\label{u12}
\end{eqnarray}
As usual, $C_A, C_F$ and $T_F$ are the group coefficients of the gauge
group  and $n_F$ is the number of flavors. \\

The results of our calculations for $\Pi_{1,2}$ read
\begin{eqnarray}
\label{pi1}
&&\Pi_1\left(\frac{Q^2}{m^2}\right)=\frac{4}{3z}[1-(1+2z)(1-z)G(z)]\;,
 \\
&&\nonumber \\
\label{pi2}
&&\Pi_2\left(\frac{Q^2}{m^2}\right) =\frac{(1+2z)}{3z^2}
  [(C_A+4C_F)~\sigma(z)-(C_A-2C_F) (1-2z)~I(z)] \nonumber \\
&&~~+\frac{2}{9z} \left\{ 39+3\tilde{I}_3^{(4)}(z)
-[4z^2+134z+57-12(2-5z) z G(z)]~(1-z) G(z) \right.  \nonumber \\
&&~~\left. ~~~~~~~~~~~~~~+2[z^2+18z+9-3(3+8z)(1-z) G(z)] \ln(-4z)
    \right\} C_A \nonumber \\
&&~~+\frac{2}{3z}\left\{13-[ 6(3+2z)
  +(7+8z-48z^2)G(z)](1-z)G(z)\right\}C_F\;,
\end{eqnarray}
where
\begin{equation}
Q^2=-q^2~,~~~~~~z=\frac{q^2}{4m^2}~,~~~~~~y=\frac{\sqrt{1-1/z}-1}
                                {\sqrt{1-1/z}+1}\;,
\end{equation}
denote the kinematic variables and
\begin{eqnarray}
&&G(z)=\frac{2y\ln y}{y^2-1}\;, \\
&& \nonumber \\
&&I(z)=6[\zeta_3+4 \Li3(-y)+2\Li3(y)]-8[2\Li2(-y)+\Li2(y)]\ln y
 \nonumber \\
&&~~~~~~~~~~~~~~~~~~~~~~~~~~~~~~~~
~              ~~~~~~~~~-2 [2\ln(1+y)+\ln(1-y)] \ln^2y\;, \\
&&\tilde{I}^{(4)}_3(z)=6\zeta_3-6 \Li3(y)+6\ln y \Li2(y)
 +2 \ln(1-y) \ln^2y\;,\\
&&\sigma(z)=\frac{1-y^2}{y} \left \{2 \Li2(-y)+\Li2(y)
 +[\ln(1-y)+2\ln(1+y)-\frac34 \ln y]\ln y \right\}
\end{eqnarray}
are our basic integrals. The functions $I(z)$, $\tilde{I}_3^{(4)}(z)$
are master integrals considered in~\cite{Brod,BFT}.

Setting $C_A=0, C_F= T_F =1, n_F=1$ and taking the limit $\mu^2 \to 0$
we reproduce the well known result for the photon propagator~\cite{BFT,KS}
in the on-shell scheme.

At large Euclidean momentum $Q^2=-q^2$ we find the asymptotic forms
\begin{eqnarray}
&&\Pi_1\left(\frac{Q^2}{m^2}\right) \stackrel{Q^2\ra \infty}{\simeq}
-\frac43 \ln \frac{Q^2}{m^2}-8\left(\frac{m^2}{Q^2}\right)
 +8 \left(\frac{m^2}{Q^2}\right)^2\ln \frac{Q^2}{m^2}
 + \cdots , \nonumber \\
&& \nonumber \\
&&\Pi_2\left(\frac{Q^2}{m^2}\right) \stackrel{Q^2\ra \infty}{\simeq}
 -\frac43(5C_A+3C_F)
  \ln \frac{Q^2}{m^2}+\frac29(C_A+36(C_A-2C_F)\zeta_3)
 \nonumber \\
&&~~~~~~~~~~~~~~~~~~~~~~~~~~~
-6(3C_A-8C_F)\left(\frac{m^2}{Q^2}\right) \ln \frac{Q^2}{m^2}
 + \cdots .
\label{ASYM}
\end{eqnarray}
With these results at hand we are able now to obtain the
mass--dependent two--loop $\beta$--function.

\section{The RG equation and the effective coupling}
In the BFM the RG $\beta$--function is given by
\begin{equation}
\mu^2 \frac{d}{d\mu^2 }\:\left(\frac{\alpha_s}{4\pi}\right)=
 \lim_{\varepsilon \to 0} ~\alpha_s \: \mu \frac{\partial}{\partial
  \mu} \ln Z_A=-\beta_0 \left( \frac{\alpha_s}{4\pi} \right)^2 -\beta_1
  \left( \frac{\alpha_s} {4\pi} \right)^3 -\cdots
\end{equation}
and hence the coefficients of the $\beta$--function may be simply
obtained by differentiating (\ref{pi1}) and (\ref{pi2}). The results read
\begin{eqnarray}
&&\beta_0=\frac{11}{3}C_A-\frac43 T_F \sum_{i=1}^{n_F}
 b_0 \left( \frac{\mu^2}{m_i^2}\right), \nonumber \\
&& \nonumber \\
&&\beta_1=\frac{34}{3}C_A^2-~~T_F \sum_{i=1}^{n_F}
 b_1\left( \frac{\mu^2}{m_i^2}\right),
\end{eqnarray}
where
\begin{eqnarray}
&&b_0\left( \frac{\mu^2}{m^2} \right)=1+\frac{3}{2x}(1-G(x))\;,\nonumber  \\
&& \nonumber \\
&&b_1\left( \frac{\mu^2}{m^2} \right)=[16(1-x^2)C_F+(1+8x^2)C_A]
 \frac{\sigma(x)}{6x^2(1-x)}-\frac{2}{3x^2}(C_A-2C_F) I(x) \nonumber \\
&&~~+\frac{2}{3x}\tilde{I}_3^{(4)}
 C_A+[(1+3x-10x^2+12x^3)C_A-3(3-3x-4x^2+8x^3)C_F]\frac{4}{3x}G^2(x)
 \nonumber \\
&&~~ -[(147-4x-100x^2+8x^3)C_A+168(1-x)C_F + 6 (9 + 4 x) \ln(-4x)C_A]
 \frac{1}{9x}G(x) \nonumber \\
&&~~+[(99+62x)C_A+12(11+3x)C_F+2(27+24x-2x^2)\ln (-4x)C_A]
 \frac{1}{9x}\;,
\end{eqnarray}
with $x=-\mu^2/(4m^2)$. This is our main result.\\

In Ref.~\cite{YH} the $\beta$--function for QCD ( $C_A=3$, $C_F=4/3$,
$T_F=1/2$) was evaluated in the standard approach with a renormalized
coupling constant defined via the gluon-ghost-ghost vertex in the
Landau gauge taken at the symmetric Euclidean point. The authors
presented only an approximate result for the function
\begin{equation}
B_1(r)=\frac{34C_A^2-3 \beta_1}{4T_F\:(5C_A+3C_F)}
\end{equation}
which corresponds to our function $b_1(r)$ and which they parametrized
as
\begin{equation}
B_1(r)=\frac{(-0.45577+0.26995r)r}{1+2.1742 r + 0.26995r^2}
\label{japan}
\end{equation}
with $ r=\mu^2/m^2$.
As asserted in~\cite{YH} the parametrization (\ref{japan}) has the
maximum deviation from the true value in the entire range $0\leq r
\leq \infty$ smaller than 0.005. We find that the difference between
our expression and (\ref{japan}) in the same region is less than
0.015, which is also very small. This is somewhat surprising, since we
are comparing couplings in different schemes.

For a mass--dependent renormalization schemes the RG equations
\bea
\mu \frac{d}{d\mu} g_s(\mu) =\beta[g_s(\mu),m_j(\mu)/\mu]\;\;,\;\;\;
\mu \frac{d}{d\mu} m_i(\mu) =-\gamma_m[g_s(\mu),m_j(\mu)/\mu]\:m_i(\mu)
\label{RGM}
\eea
in general can be solved only by numerical integration. However,
an approximate solution for the mass dependent effective QCD
coupling was proposed in~\cite{DV1,DV2}. Indeed, at the two-loop
level the expression
\begin{equation}
\overline{\alpha_s}(Q^2)=\frac{\alpha_s}
 {1+\alpha_s/(4\pi) U_1+\alpha_s/(4\pi) (U_2/U_1)
 \ln(1+\alpha_s/(4\pi) U_1)}\;,
\label{alpha}
\end{equation}
with $U_{1,2}$ given in (\ref{u12}),
correctly sums up all leading as well as ``next-to-leading''
terms $\alpha_s U_2 (\alpha_s U_1)^n$ though it is not an exact solution
of the two-loop differential RG equation.
We will compare (\ref{alpha}) with the result of the numerical
integration of the RG equation below.

\section{\BFMOM coupling in terms of the \MSb coupling}
Let us define the auxiliary functions
\bea
&&z_{1i}=-\Pi_1 \left( r_i \right)
 - \frac{20}{9}
 -\frac43  ~l_i \;, \nonumber \\
&& \nonumber \\
&&z_{2i}=- \Pi_2 \left(r_i \right)
 - \left( \frac{52}{3}+\frac{20}{3} l_i \right) C_A
 - \left(\frac{55}{3} +4 l_i \right) C_F\;,
\eea
where
\bea
l_i=\ln r_i\;,~~~~~~r_i=\frac{\mu^2}{m_i^2}\;. \nn
\eea
For later use we note that for light fermions, utilizing the
expansion~(\ref{ASYM}), we obtain
\bea
&&z_{1i}=- \frac{20}{9} + O(m_i^2/\mu^2) \;, \nonumber \\
&& \nonumber \\
&&z_{2i}=
 - \left(\frac{158}{9}+8\:\zeta_3 \right) C_A
 - \left(\frac{55}{3} -16 \: \zeta_3 \right) C_F+ O(m_i^2/\mu^2)\;.
\label{CONST}
\eea
The relationship between the renormalized coupling constants may then
be written in the form
\bea
\bar{h}&=&\frac{\alpha_{s\: \mathrm{MOM}}}{4 \pi}=H(h,\mu^2)
 =h+k_1(\mu^2)h^2+(k_2(\mu^2)+k^2_1(\mu^2)) h^3+\cdots
\label{RELATION}
\eea
where
\bea
&&k_1(\mu^2)= \frac{205}{36}C_A+ T_F \sum_{i=1}^{n_F} z_{1i} \;,
 \nonumber \\
&& \nonumber \\
&&k_2(\mu^2)=\left(\frac{2687}{72} -\frac{57}{8}\zeta_3\right)
  C_A^2 +T_F \sum_{i=1}^{n_F} z_{2i}\;,
\label{KS}
\eea
and
\bea
h \equiv h_{\MSbm}&=&\frac{\alpha_{s\: \MSbm}}{4 \pi}\;. \nn
\eea
Differentiating the relation with respect to  $\mu^2$
we obtain:
\bea
\mu^2 \frac{d \bar{h}}{d \mu^2}
&=& \beta_{\mathrm{MOM}}(\bar{h})
 = \frac{\partial H(h,\mu^2)}
 {\partial h} \beta(h)+\mu^2 \frac{\partial H(h,\mu^2)}
 {\partial \mu^2}
\crn && \crn
&=&-\beta_{0\:\mathrm{MOM}}\:\bar{h}^2
 -\beta_{1\:\mathrm{MOM}}\:\bar{h}^3-\cdots \;,
\eea
where $\beta(h)$ is the $\beta$ function in the \MSb
scheme:
\bea
\beta(h)=-\beta_0h^2-\beta_1h^3-\cdots\;,
\eea
with
\bea
&&\beta_0=\frac{11}{3}C_A-\frac{4}{3} T_F n_F\;, \nonumber \\
&& \nonumber \\
&&\beta_1=\frac{34}{3}C_A^2-\frac{20}{3}C_AT_Fn_F- 4 C_F T_F n_F\;.
\eea
From the above equation we obtain
\bea
&&\beta_{0\: \mathrm{MOM}}=\beta_0
 -\mu^2 \frac{\partial}{\partial \mu^2}k_1(\mu^2), \nonumber \\
&& \nonumber \\
&&\beta_{1\: \mathrm{MOM}}=\beta_1
 -\mu^2 \frac{\partial}{\partial \mu^2} k_2(\mu^2).
\eea
As it should be, in the massless limit the $\beta$--functions
agree.


\section{\BFMOM versus \MSb coupling: numerical aspects}
For numerical studies we use the following pole quark masses~\ci{PDG98}
$$m_u \sim m_d \sim m_s \sim 0\;;\;\;m_c=1.55 \gv \;;\;\;m_b=4.70 \gv
\;;\;\;m_t=173.80 \gv \;.$$
For the strong interaction coupling we take $\alpha_{s\; {\small
\MSbm}}^{(5)}$ = 0.12$\pm0.003$ at scale $M_Z$=91.19 GeV~\ci{LEP}.  
In Fig.~1 we show that Shirkov's formula (\ref{alpha}) provides an
excellent approximation to the exact solution of the two--loop RG
equation. At sufficiently large scales the mass effects in the
$\beta$--function are small and we expect no large numerical
differences between different schemes. This is illustrated in Fig.~2,
where the evolution of the running couplings is shown for a common
start value of $\alpha_s$ = 0.12 at the scale $M_Z$ = 91.19 GeV. Only
the space--like $E=\sqrt{-q^2}$ is considered. We see that the mass
effects are of comparable size as the 3--loop
contribution~\ci{fourloop,matching} in the \MSb scheme (see Tab.~1
given below). The \MSb results were obtained by adopting the
Bernreuther--Wetzel~(BW)~\ci{DecouplingBW} matching scheme between the
effective theories with different flavors. We checked that utilizing
Marciano~(M) matching~\ci{DecouplingM}, instead, leads to answers
somewhat closer to our \MOM results. Only the latter one exhibit the
correct physical mass behavior.

\rput{90}(9.0,-3.0){\scalebox{.7 .7}{%
\epsfbox{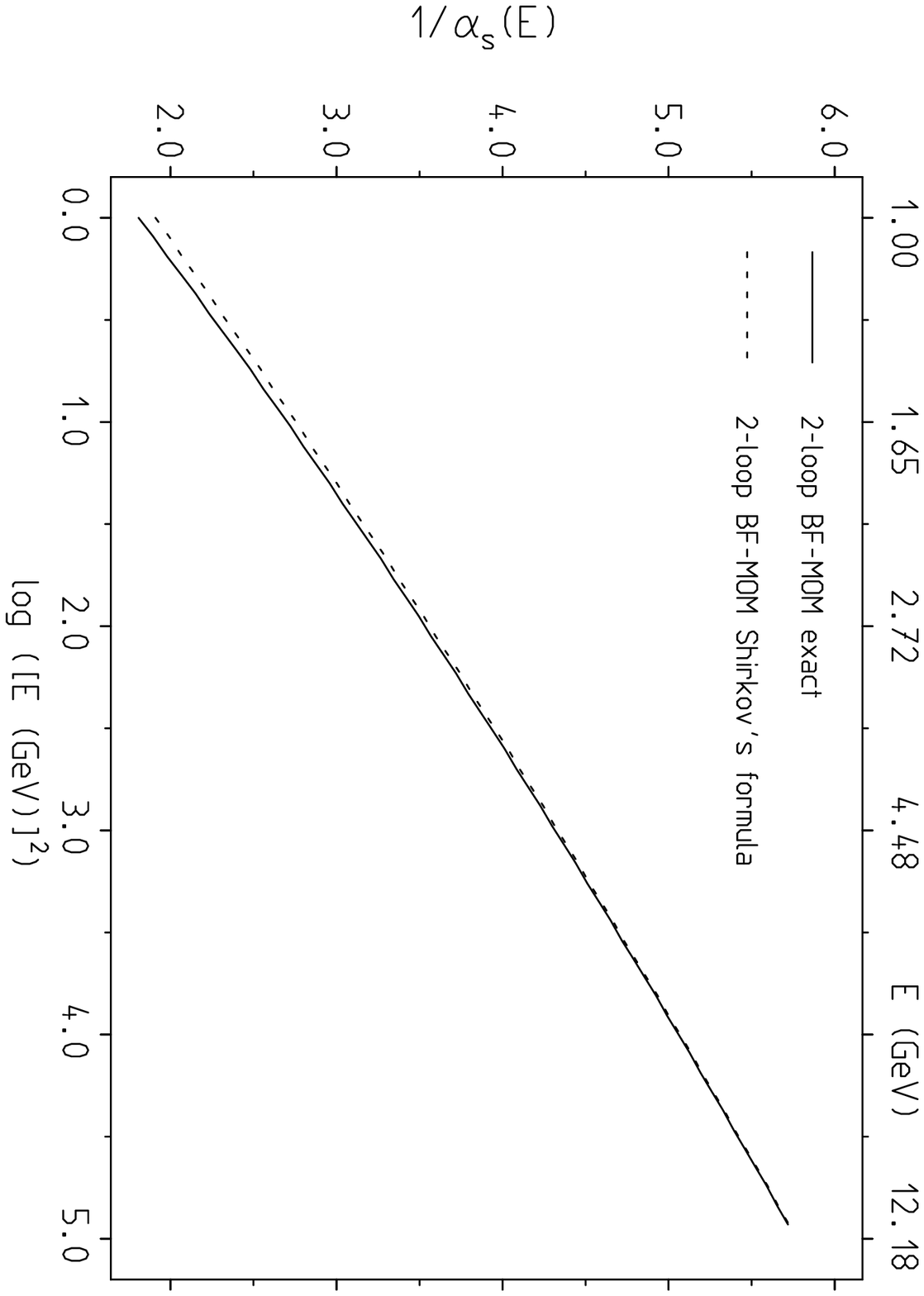}}}

\vspace*{8.5cm}

\begin{center}
\begin{minipage}[h]{15.2cm} \baselineskip 12truept \noi
\small{{Figure~1:
  Evolution of $\alpha_s$ in the \BFMOM scheme
  normalized to $\alpha_s$ = 0.12 at the scale $M_Z$ =
  91.19 GeV. The dotted line represents the approximation by
 Shirkov's formula.
}}
\end{minipage}
\end{center}


\rput{90}(9.0,-4.0){\scalebox{.7 .7}{%
\epsfbox{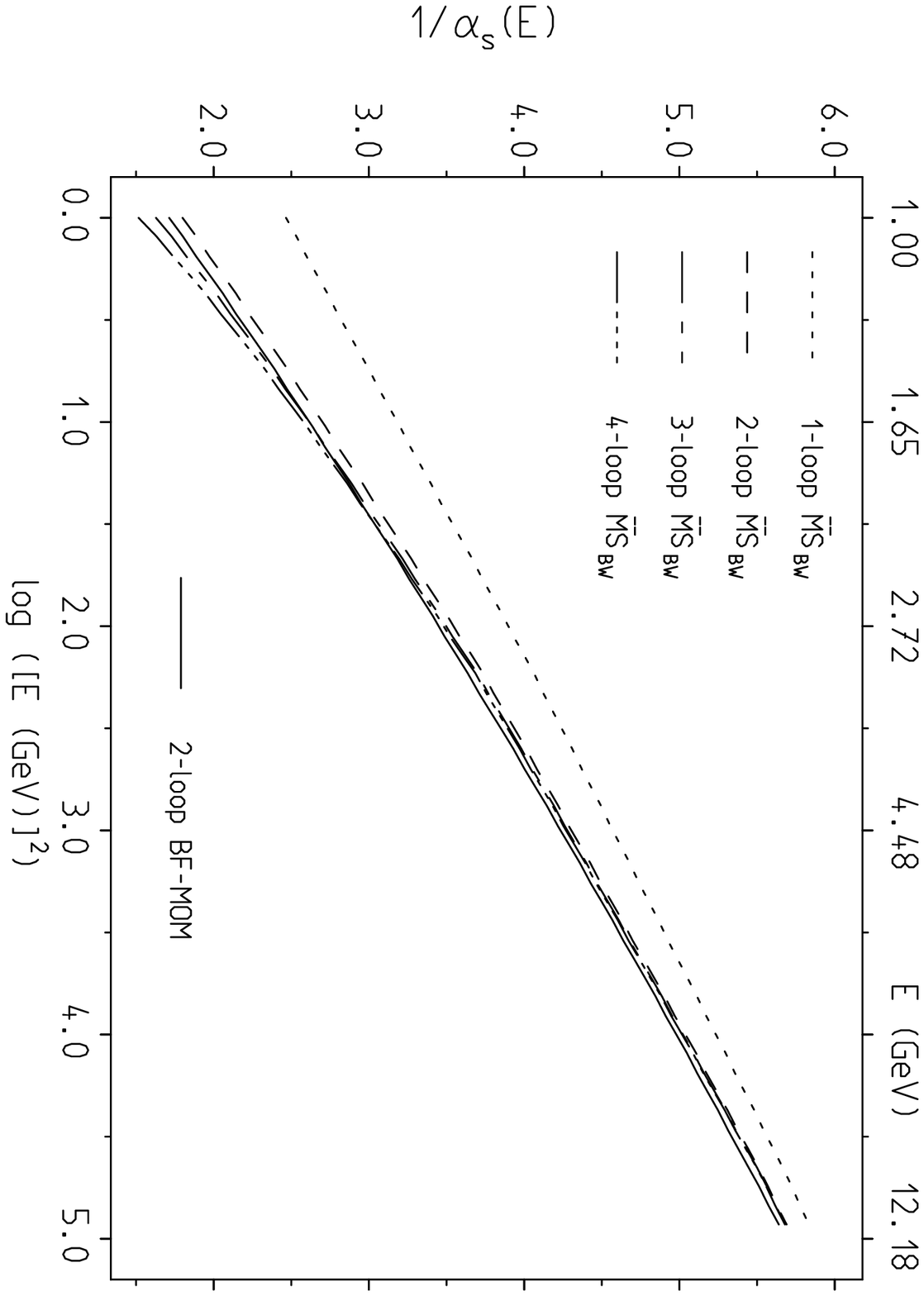}}}

\vspace*{10cm}

\begin{center}
\begin{minipage}[h]{15.2cm} \baselineskip 12truept \noi
\small{{Figure~2:
  Comparison of the $\alpha_s$ evolution in the space--like region
  normalized to a common value $\alpha_s$ = 0.12 at scale $M_Z$ =
  91.19 GeV. The dotted, dashed, dash--dot and the dash--dot--dot--dot
  curves show, respectively, the one--loop, two--loop, three--loop and
  the four--loop \MSb evolution for BW--matching. The full line
  represents the exact \BFMOM running coupling.}}
\end{minipage}
\end{center}

\vspace*{5mm}

Although BW--matching seems to be better justified from a field
theoretical point of view, it leads to ``threshold jumps'' which of
course are not physical in the space--like region. In contrast
M--matching assumes continuity of $\alpha_s$ across the matching scale
(``thresholds'').\\ 

While there are no really large numerical
differences in the $\beta$--functions, i.e., the derivatives of
$\alpha_s$ with respect to $\mu$, down to moderately low scales, there
are large $\mu$-- and hence mass--independent terms in the
relationship between the coupling constants~(\ref{RELATION}), as
follows from ~(\ref{KS}) and~(\ref{CONST}), and as we can see in
Fig.~3.  A large constant shift in

\rput{90}(9.0,-4.0){\scalebox{.7 .7}{%
\epsfbox{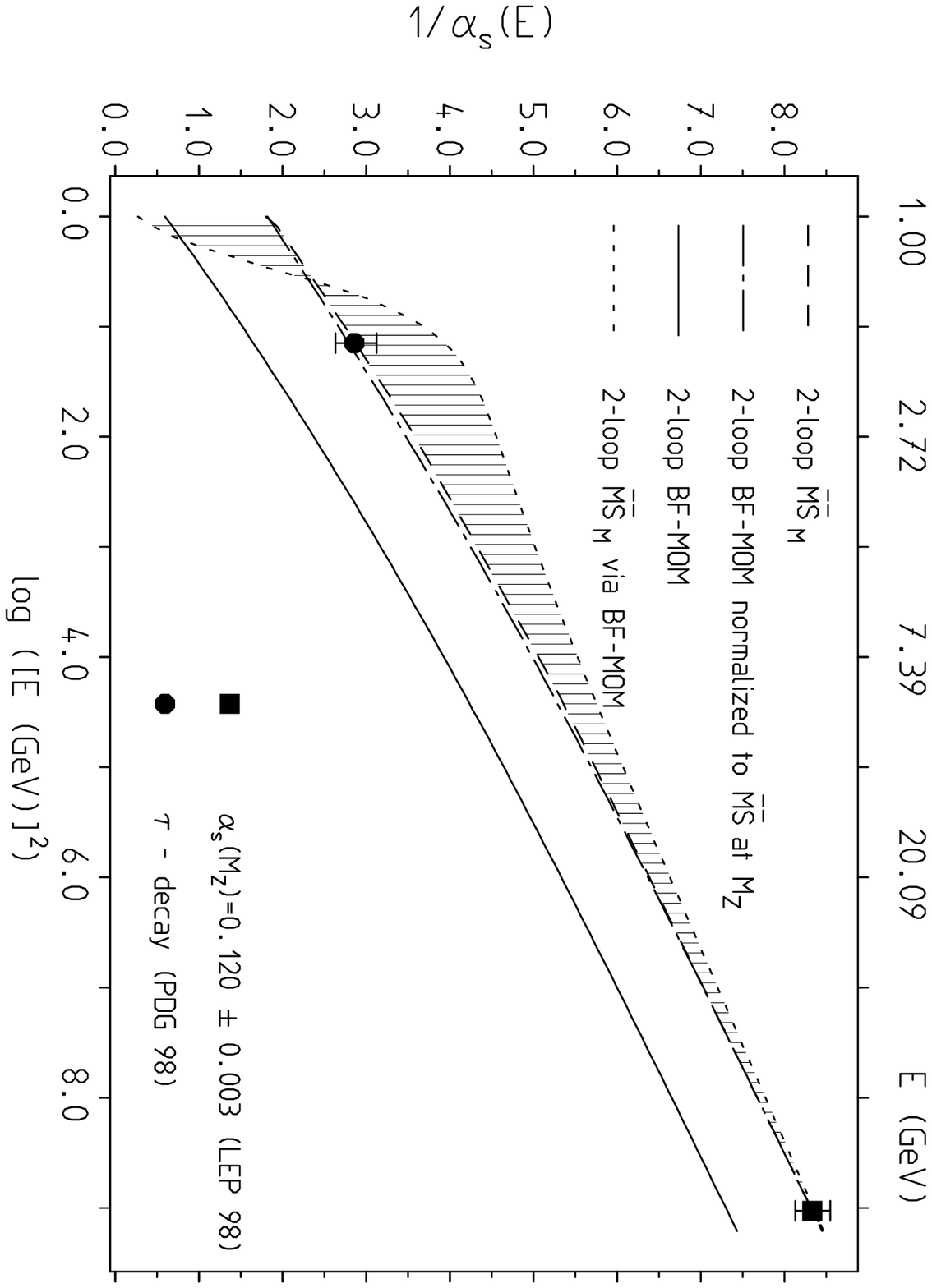}}}

\vspace*{10cm}

\begin{center}
\begin{minipage}[h]{15.2cm} \baselineskip 12truept \noi
\small{{Figure~3:
  Comparison of $\alpha_s$ in the \BFMOM and the \MSb schemes with
  $\alpha_{s\; {\small \MSbm}}^{(5)}$ = 0.12 at scale $M_Z$=91.19 GeV
  and $\alpha_{s\; {\small \BFMOMm}}$ at this scale calculated using
  (\ref{RELATION}).  The dotted line is the \MSb coupling calculated
  from $\alpha_{s\;{\small \BFMOMm}}(E)$ by inverting
  (\ref{RELATION}).}}
\end{minipage}
\end{center}

\vspace*{6mm}

$\alpha_s$ of
about plus 14\% at $M_Z$ is obtained when we go from the \MSb to the
\MOM scheme. In principle, this does not affect the prediction of
physical observables. However, the scheme dependence which is due to
truncation errors of the perturbation expansion is different for
different renormalization schemes. The shaded area of Fig.~3 reflects
the theoretical uncertainty at the two--loop level which shows up in
the comparison of the two schemes. Below about 1.15(2.92) GeV the
one--loop correction $k_1(\mu^2)\:h$ in (\ref{RELATION}) exceeds the
leading trivial term by 100(50)\% and the perturbation expansion
cannot be applied any longer (see Fig.~3).\\

The occurrence of the disturbing large numerical constants in the
relationship between the renormalized couplings belonging to different
renormalization schemes is not a peculiar feature in the relation
between the \MOM and \MSb schemes.  Similar worrying lare terms, long
time ago, were the reason for replacing the original \MS by the \MSb
scheme~\ci{MSb}, which are related by a simple rescaling of the scale
parameter $\mu$. Other, more sophisticated, examples of eliminating
leading terms by rescaling were proposed in Ref.~\ci{BLM}. Also, for
the comparison of non-perturbative calculations of running couplings
in lattice QCD with perturbative results, the adequate choice of a
relative scale factor turns out to be crucial~\ci{RSSchladming}. The
rescaling usually leads to dramatically improved agreement. A
condition for the rescaling to make sense is that the
$\beta$--functions of the two schemes under consideration do not
differ too much numerically. For our two schemes this condition is
fairly well satisfied (see Fig.~2). In fact at higher energies the
$\beta$--functions become identical. Such a rescaling procedure thus
looks natural if we tune the running couplings to agree with good
accuracy at high energies. This can be achieved as follows: While
(\ref{RELATION}) reads ($\tilde{k}_2=k_2+k_1^2$)
\bea
\bar{h}(\mu^2)=h(\mu^2)+k_1 h^2(\mu^2)+\tilde{k}_2 h^3(\mu^2)+ O(h^4)
\eea
we may absorb the disturbing large term $k_1$ into a rescaling of
 $\mu$ by a factor $x_0$ such that~\ci{RSSchladming}
\bea
\bar{h}((x_0\mu)^2)=h(\mu^2)+0+ O(h^3)\;.
\eea
Expanding the RG solution (\ref{alpha}) we have ($\tilde{U}_2=U_2-U^2_1$)
\bea 
\bar{h}((x_0\mu)^2) &=& \bar{h}(\mu^2)
-U_1(x^2_0,\{m_i^2/\mu^2\})\bar{h}^2(\mu^2)
-\tilde{U}_2(x^2_0,\{m_i^2/\mu^2\})\bar{h}^3(\mu^2)+
O(\bar{h}^4) \nn \\
&=& h(\mu^2)+\left(k_1-U_1(x^2_0,\{m_i^2/\mu^2\})\right)\:h^2(\mu^2) \\&&
+\left(\tilde{k}_2-2U_1(x^2_0,\{m_i^2/\mu^2\})k_1 
-\tilde{U}_2(x^2_0,\{m_i^2/\mu^2\})\right)\:h^3(\mu^2) 
+ O(h^4)\nn
\label{MOMMSbr}
\eea 
and the rescaling factor $x_0$ is determined by the equation
\bea
k_1=U_1(x^2_0,\{m_i^2/\mu^2\})\;.
\eea
In our mass dependent scheme we require this to be true only at very
large scales $\mu^2 \gg m_f^2 $ for all flavors $f$ including the top
quark. This convention is simple and most importantly, it does not
conflict with the manifest decoupling property of the MOM scheme. As a
consequence we obtain a running coupling which depends very little on
the scheme at large energies, a property which looks most natural in
an asymptotically free theory like QCD. For the \BFMOM scheme the
rescaling factor $x_0$ is determined by
\bea
\ln (x^2_0)=\left(\frac{205}{36}C_A-\frac{20}{9}T_F n_F \right)/
\left(\frac{11}{3}C_A-\frac{4}{3}T_F n_F \right)=125/84
\eea
for QCD with $n_F=6$ flavors. Numerically we find $x_0\simeq
2.0144$.\\

In order to check whether the above rescaling makes sense, we
must inspect the change of the 2--loop coefficient in the rescaled
relationship (\ref{MOMMSbr}) between \MOM and \MSb. Indeed, the
rescaling changes the coefficients from $k_1\simeq 10.42,\;\;
\tilde{k}_2\simeq 126.35$ to $k_{1\:\mathrm{eff}}= 0,\;\;
\tilde{k}_{2\:\mathrm{eff}}\simeq -32.46$ and thus we get a
substantial improvement for the next to leading coefficient too, as it
should be. We note that the rescaling improved \MOM perturbation
expansion at low energies does not any longer deviate substantially
from the \MSb results. Of course, only the appropriate higher order
calculations of observables in the \BFMOM scheme could reveal the true
convergence properties of the perturbation series in this scheme.\\

In the \MOM scheme the energy scale comes in by a momentum subtraction
and the location of the thresholds of course cannot depend on the
rescaling ``reparametrization''. This means that actually the scale
must be changed in the \MSb scheme, where the scale parameter $\mu$
enters in a purely formal way and ``thresholds'' are put in by hand
for switching between the effective theories of different numbers of
flavors. Since, conventionally, $\mu$ in the \MSb scheme has already
been identified with the c.m. energy, for example, in the LEP
determination of $\alpha_s(M_Z)$ which we use as an input, we have to
apply the rescaling to the \MOM calculation. As the thresholds must
stay at their ``physical'' location, i.e., $4m^2/q^2$ must remain
invariant, we have to perform the scaling simultaneously to the energy
and the masses.\\

The result from utilizing this rescaling procedure is displayed in
Fig.~4.  The large deviations seen in Fig.~3 have disappeared now. 
The sizes of effects are still illustrated by what we observe in
Fig.~2 except that the initial values at $M_Z$ differ.  
In Fig.~4 we have recalculated the input values of
$\alpha^{(5)}_s(M_Z)$ as a function of the perturbative order,
assuming the observable $R(s)$ to have a given experimental value.
$R(s)$ is the ratio of hadronic to leptonic $e^+e^-$--annihilation
cross sections at sufficiently large $s$, from which a precise
determinations of $\alpha_s(s)$ is possible. At our reference scale
$M_Z$ we may use perturbative QCD in the massless approximation
\ci{Rpert3}
\bea
R(s)&=&3\sum_f Q_f^2\:\left(1+a+c_1 a^2+c_2 a^3 +\cdots \right)
\label{Rpert}
\eea
where $Q_f$ denotes the charge of the quark, $a=4h=\alpha_s (s)/\pi$, and
\bea
c_1&=&~~1.9857-0.1153\: n_F \crn c_2&=&-6.6368-1.2002\: n_F - 0.0052\: n_F^2
-1.2395\: (\sum Q_f)^2/(3 \sum Q_f^2)\nn
\eea
in the $\overline{MS}$ scheme, with $n_F=5$ active flavors.\\

Some concluding remarks: We have investigated a \MOM renormalization
scheme in the background field gauge at the two--loop order in QCD and
shown that a substantial scheme dependence is observed relative to the
\MSb scheme, unless we apply a suitable rescaling. These findings are 
in accord with earlier investigations at the one--loop~\ci{GP,CG79}
and two--loop~\ci{YH} level. Mass effects in any case are
non-negligible at a level of precision where also higher order
corrections are relevant. The calculation in full QCD includes the
exact mass effects and is smooth and analytic at all scales and in
particular across thresholds. It thus avoids problems with the \MSb
scheme addressed in a recent article by Brodsky et al.~\ci{Brodsky98}
which were cured by an analytic extension of the
\MSb renormalization scheme.\\

We note that the use of the \BFMOM scheme, particularly when using the
compact form obtained for Shirkov's approximation, is much easier in
practice because decoupling is manifest at any threshold and there are
no matching conditions to be imposed.\\

\rput{90}(9.0,-4.0){\scalebox{.7 .7}{%
\epsfbox{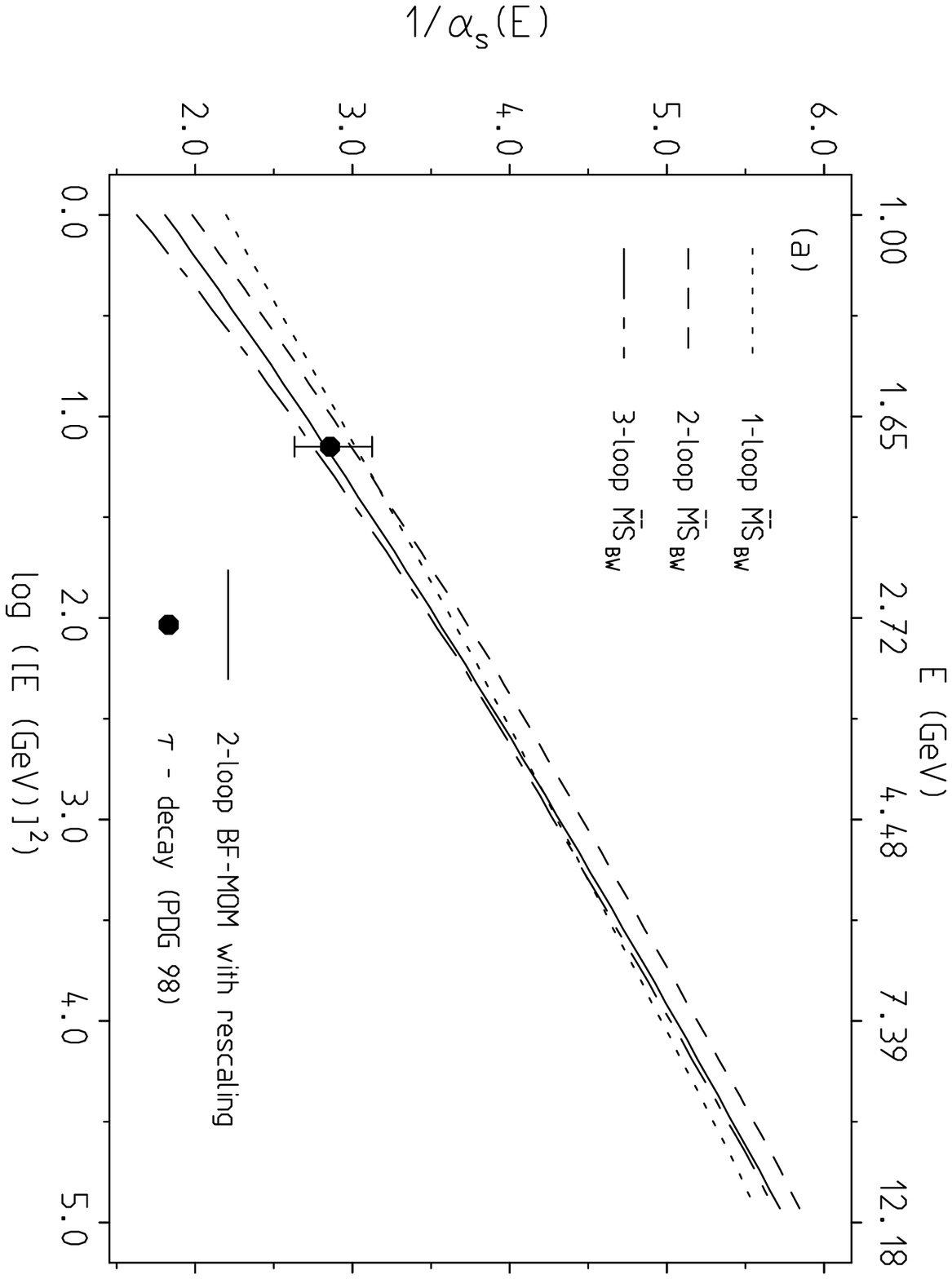}}}

\vspace*{10cm}

\rput{90}(9.0,-4.0){\scalebox{.7 .7}{%
\epsfbox{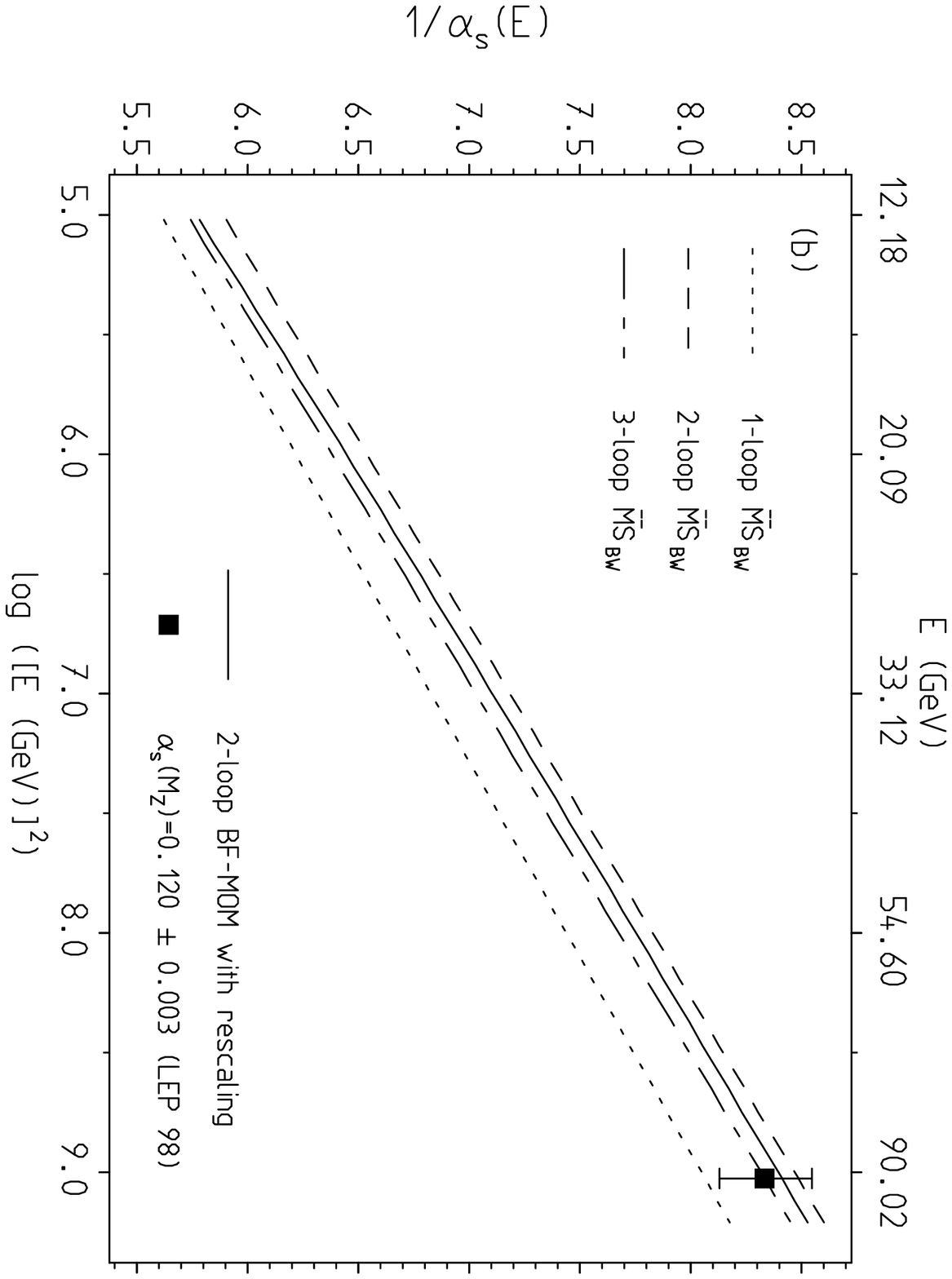}}}

\vspace*{10cm}

\begin{center}
\begin{minipage}[h]{15.2cm} \baselineskip 12truept \noi
\small{{Figure~4:
  Comparison of $\alpha_s ((x_0E)^2)$ in the \BFMOM and $\alpha_s
  (E^2)$ in the \MSb scheme with input values $\alpha_{s\; {\small
  \MSbm}}^{(5)}$ = 0.120 at scale $M_Z$=91.19 GeV and $\alpha_{s\;
  {\small \BFMOMm}}$ = 0.1189 obtained for the rescaled energy
  $x_0M_Z\simeq$190.90 GeV ($x_0\simeq 2.1044$).}}
\end{minipage}
\end{center}

\vspace*{6mm}

We emphasize that the scheme and scale dependence of perturbativ QCD
predictions is not a matter of the order of perturbation theory alone
but may depend substantially on other details like the kind of
matching condition applied in the mass independent \MSb schemes or the
threshold and mass effects in \MOM schemes. The following table (Tab.~1)
may illustrate the kind of uncertainties we expect to encounter. We
find that 

\begin{center}
\begin{minipage}[h]{15.2cm} \baselineskip 12truept \noi
\small{{Table~1:
  Comparison of predicted $\alpha_s$ values at the masses of the $\Upsilon$,
  ${\mathrm{J}/\psi}$ and $\tau$. Here we adopt a common input value 
  $\alpha_s(M_Z)=0.12$ for
  the \MSb scheme independent of the perturbative order.\\}}
\end{minipage}

\begin{tabular}{lcccl}
\hline
scheme  & $\alpha_s(M_Z)$  (input) & $\alpha_s(M_\Upsilon)$
& $\alpha_s(M_{\mathrm{J}/\psi})$& $\alpha_s(M_\tau)$ \\
\hline
\hline
~\MSb 2-loop (BW) & 0.120 &0.179 &0.260 & 0.354 \\
\hline 
~\MSb 3-loop (BW) & 0.120 &0.179 &0.262 & 0.364 \\
\hline 
~\MSb 4-loop (BW) & 0.120 &0.179 &0.263 & 0.368 \\
\hline
~\MSb 2-loop (M) & 0.120 &0.179 &0.258 & 0.348 \\
\hline
~\MSb 2-loop via \BFMOM & 0.120 &0.168 &0.211 & 0.254 \\
\hline 
(\BFMOM 2-loop    & 0.120 &0.180 &0.265 & 0.358) \\
\hline 
~\BFMOM 2-loop    & 0.137 &0.222 &0.372 & 0.605 \\
\hline 
~\BFMOM 2-loop rescaled   & 0.121 &0.181 &0.260 & 0.345  \\
\hline 
\end{tabular}
\end{center}
the 2--loop \MSb value at the $\tau$--mass $M_\tau$ is
$\alpha_s$ = 0.254 when we switch from \MSb to \BFMOM at the $Z$--mass
$M_Z$ use \BFMOM evolution down to $M_\tau$ and switch back from
\BFMOM to \MSb. Standard (direct) \MSb evolution depends on the
matching scheme utilized (BW or M) and for BW(M)--matching yields
$\alpha_s$ = 0.354(0.348), such that via \BFMOM we get a value which
is lower by 0.100(0.094). However, the \BFMOM value obtained with the
rescaling is 0.345, not very different from its \MSb value. The
Particle Data Group~\ci{PDG98} quotes $\alpha_s(M_\tau)=0.35\pm0.03$
for the experimental value obtained from $\tau$--decays 
(see also~\ci{seealso}).

\bigskip

Experience with many physical applications of the \MSb scheme somehow
established this scheme as a preferred one, in the spirit that this
prescription is better than others in the sense that it leads to
reliable perturbative predictions for many physical observables. In
our opinion it remains unclear whether a prefered scheme exist. The
problem is the appropriate choice of scale. We advocate here to take
more serious the physical mass dependence. In order to get a better
understanding of the scheme dependences we need more calculations in
different schemes. \\

{\large \bf Acknowledgments}\\

We gratefully acknowledge helpful discussions with S. Brodsky, W. van
Neerven and R. Sommer. In addition, we thank R. Sommer for carefully
reading the manuscript.

\newpage

{\Large \bf Appendix A: Bare BF propagator}
\small
\begin{eqnarray}
&&(d-1)(d-4)U_{2\:{\rm bare}}=
\frac{c_{1}}{16(d-4)(d-6)}  C_A^2 J_{111}(0,0,0)
+\frac{q^2c_2}{64}C_A^2  G_{11}^2(0,0)\nonumber \\
&&+T_F \sum_{i=1}^{n_F} \left\{ 
32 z_i m_i^4(d-4)~C_A~ \tilde{I}_3^{(d)}(z_i)
 +\frac{4 f_1(z_i)}{1-z_i}  (C_A-2C_F)~ m_i^2 G_{11}^2(m_i^2,m_i^2)
\right. \nonumber \\
&&~~+2(d-4) f_2(z_i)~ C_A~m_i^2 G_{11}(0,0) G_{11}(m_i^2,m_i^2)
\nonumber \\
&&~~+\frac{4}{1-z_i} \left[ \frac{d-2}{d-3} f_3(z_i) C_F+2f_4(z_i)C_A \right]
 G_{11}(m_i^2,m_i^2) G_{10}(m_i^2,0) \nonumber \\
&&~~ +\frac{f_5(z_i)}{15}(d-2)(d-4)~C_A~G_{11}(0,0) G_{10}(m_i^2,0)
 \nonumber \\
&&~~+\left[\frac{4 f_6(z_i)}{d-3}  C_F + \frac{f_7(z_i) C_A}{15 z_i(1-z_i)}
  \right] 4 m_i^2 J_{112}(0,m_i^2,m_i^2)  \nonumber \\
&&~~+\left[ 4(d-1)(d-2)(d-4) C_F + \frac{f_8(z_i) C_A}{15 z_i(1-z_i)}  \right ]
 ~ J_{111}(0,m_i^2,m_i^2)  \\
&&~~\left. -\left[ \frac{f_{9}(z_i)C_A}{30z_i(d-5)}  +2 ~f_{10}(z_i)
       C_F \right]\frac{(d-2)}{(1-z_i)(d-3)m_i^2} ~
       G^2_{10}(m_i^2,0)\right\}
. \nonumber
\end{eqnarray}

\begin{eqnarray}
&&c_{1}=(3d-8)\xi \left[-(d-1)(d^2-9d+22)(d-4)^2 \xi^2
 -(15d^5-256d^4+1685d^3 \right.\nonumber \\
&&~~ \left.-5292d^2+7744d-3960)\xi
 +33d^5-542d^4+3311d^3-9378d^2+12448d-6608 \right]
\nonumber \\
&&~~ +27112d-8128-33312d^2-51d^6+916d^5+20016d^3-6169d^4,
\nonumber \\
&& \nonumber \\
&&c_2=(d-1)(d-4)^2[(d-4)\xi+6(2d-7)]\xi^3
   +2(11d^4-144d^3+677 d^2-1296d+743)\xi^2 \nonumber \\
&&~~
-(84d^4-854d^3+2994d^2-4304d+2384)\xi
 +49d^4-403d^3+1106d^2-1392d+664, \nonumber \\
&& \nonumber \\
&&f_1=(d-2)[(d^2-7d+16)z-(d-5)(d-4)]z-2, \nonumber \\
&& \nonumber \\
&&f_2=(d-2)[(d-4)\xi^2-7d+12]z
 +(d-4)\xi^2+2((d-2)z+4)(3d-10)\xi-7d+16, \nonumber \\
&& \nonumber \\
&&f_3=2(d-2)(d^2-5d+8)z^2+(d-1)(d-3)(d-4)^2 z+d^3-6d^2+5d+8,
\nonumber \\
&& \nonumber \\
&&f_4=(d-2)(1-2z)[(d-2)z+1], \nonumber\\
&& \nonumber \\
&&f_5=4 [(d-1)\xi+3d-7](d-4)z^2-10[(d-1)(3d-8)\xi
 -7d^2+23d-28]z \nonumber \\
&&~~~-15(d-4)\xi^2-30(3d-10)\xi+15(7d-16) \nonumber \\
&& \nonumber \\
&&f_6=(d-2)[(d^2-5d+8)z-d^2+7d-10], \nonumber \\
&& \nonumber \\
&&f_7=2(d-4)^2[(d-1)\xi+3d-7]z^4
 -\left[3(7d-20)(d-1)(d-4)\xi-172d-17d^3+81d^2\right.
 \nonumber \\
&&~~\left.+240\right]z^3-[(637d^2-2331 d-54 d^3+2708)\xi
 -713d^2+2025d+78d^3-1924]z^2 \nonumber \\
&&~~ -[(d-6)(53d^2-398d+729)\xi+2202-81d^3+800 d^2
 -2447 d] z \nonumber \\
&&~~ +(2d-7)(d-7)(9d-41)\xi-26d^3+275d^2
 -892d+787, \nonumber \\
&& \nonumber \\
&&f_8=2(d-3)(d-4)(3d-8)[(d-1)\xi+3d-7]z^3 \nonumber \\
&&~~-[(3d-8)(19d^3-214d^2+715d-712)\xi
 -3891d^2+7392d-5248+880d^3-69d^4]z^2
\nonumber \\
&&~~+(d-4)[(5d-17)(7d-39)(3d-8)\xi
 -165d^3+1424d^2-3819d+3192]z\nonumber \\
&&~~-(d-7)(9d-41)(2d-7)(3d-8)\xi
 +(3d-8)(26d^3-275d^2+892d-787), \nonumber \\
&& \nonumber \\
&&f_{9}=2(d-5)(d-3)(d-4)^2[(d-1)\xi+3d-7]z^3 \nonumber \\
&&~~ -\left[(d-1)(d-4)(19d^3-208d^2+711d-754)\xi
 -6266d+345d^4+2888-23d^5 \right. \nonumber \\
&&~~ \left.+5095 d^2-1943d^3\right)z^2+(d-2)
\left[(d-6)(35d^3-401d^2+1521d-1923)\xi \right.\nonumber \\
&&~~\left.-(55d^4-817d^3+4375d^2-9799d+7434) \right] z
 - \left[(d-7)(2d-7)(9d-41)\xi \right.\nonumber \\
&&~~\left. -(26d^3-275d^2+892d-787) \right] (d-2)(d-5), \nonumber \\
&& \nonumber \\
&&f_{10}=(d-2)[(d^2-5d+8)z+d^3-7d^2+16d-14].
\end{eqnarray}
\normalsize

In the above formulae we have used the following notation:
\bea
&&\tilde{I}^{(d)}_3=\int \int \frac{d^dk_1~d^dk_2}
{\pi^d~k_1^2(k_2^2-m^2)(k_1-q)^2((k_2-q)^2-m^2)((k_1-k_2)^2-m^2)}\;,
   \non \\
&& \non \\
&&J_{\alpha \beta \gamma}(m_1^2,m_2^2,m_3^2)=
\int \int \frac{d^dk_1~d^dk_2}{\pi^d~
         (k_1^2-m_1^2)^{\alpha}
        ((k_2-q)^2-m_2^2)^{\beta}
	((k_1-k_2)^2-m_3^2)^{\gamma}}\;, \non \\
&& \non \\
&&G_{\alpha \beta}(m_1^2,m_2^2)=\int \frac{d^dk_1}{\pi^{d/2}}\frac{1}
{(k_1^2-m_1^2)^{\alpha} ((k_1-q)^2-m_2^2)^{\beta}}\;.
\eea
All parameters are the bare one's, $z_i=q^2/(4m_i^2)$ and the
coefficient functions $f_n=f_n(z_i)$ are functions of $z_i$.

\newpage

\vspace*{1.5cm}

{\Large \bf Appendix B: The BF Feynman rules.}\\

In addition to the conventional QCD Feynman rules we have:\\

\begin{picture}(40,40)(0,0)
\SetWidth{1.2}
\Photon(30,0)(30,15){1}{4}
\Photon(30,0)(43,-7.5){1}{4}
\Photon(30,0)(17,-7.5){1}{4}
\Vertex(30,0){1.8}
\Text(30,31)[]{a,$\mu$}
\Text(17,-13)[]{b,$\nu$}
\Text(44,-13)[]{c,$\lambda$}
\Text(36,7.5)[]{p}
\Text(45,0)[]{r}
\Text(15,0)[]{q}
\Text(65,4)[lb]{$gf_{abc} \left[ g_{\mu \lambda}\:(p-r-{\frac{1}{\xi}} q)_\nu +
g_{\nu \lambda}\:(r-q)_\mu \right.$}
\Text(80,-15)[lb]{$ \left. +g_{\mu \nu}\:(q-p+{\frac{1}{\xi}}
r)_\lambda \right]$}
\end{picture}
\cput[]{00}(-0.42,0.7){\scalebox{0.5}{~~}}
\rput{00}(-0.40,0.72){A}
\rput{00}(11.5,0.0){($\xi=\infty$ standard triple vertex)}

\vspace*{1.2cm}

\begin{picture}(40,40)(0,0)
\SetWidth{1.2}
\Photon(30,0)(15, 15){1}{5}
\Photon(30,0)(45, 15){1}{5}
\Photon(30,0)(45,-15){1}{5}
\Photon(30,0)(15,-15){1}{5}
\Vertex(30,0){1.8}
\Text(15,30)[]{a,$\mu$}
\Text(15,-20)[]{b,$\nu$}
\Text(45,-20)[]{c,$\lambda$}
\Text(45,+20)[]{d,$\rho$}
\Text(65,14)[lb]{$-ig^2 \left[ f_{abx}f_{xcd}
(g_{\mu \lambda}g_{\nu \rho}-g_{\mu \rho}g_{\nu \lambda})
\right.$}
\Text(80,-6)[lb]{$ \left. +f_{adx}f_{xbc}
(g_{\mu \nu}g_{\lambda \rho}-g_{\mu \lambda}g_{\nu \rho})
\right.$}
\Text(80,-25)[lb]{$ \left. +f_{acx}f_{xbd}
(g_{\mu \nu}g_{\lambda \rho}-g_{\mu \rho}g_{\nu \lambda})
\right]$}
\end{picture}
\cput[]{00}(-1.01,0.65){\scalebox{0.5}{~~}}
\rput{00}(-0.99,0.67){A}
\rput{00}(11.5,0.0){( = standard quartic vertex)}

\vspace*{1.2cm}

\begin{picture}(40,40)(0,0)
\SetWidth{1.2}
\Photon(30,0)(15, 15){1}{5}
\Photon(30,0)(45,-15){1}{5}
\Photon(30,0)(45, 15){1}{5}
\Photon(30,0)(15,-15){1}{5}
\Vertex(30,0){1.8}
\Text(15,30)[]{a,$\mu$}
\Text(15,-20)[]{b,$\nu$}
\Text(47,-30)[]{c,$\lambda$}
\Text(45,+20)[]{d,$\rho$}
\Text(65,14)[lb]{$-ig^2 \left[ f_{abx}f_{xcd}
(g_{\mu \lambda}g_{\nu \rho}-g_{\mu \rho}g_{\nu \lambda}
+{\frac{1}{\xi}}g_{\mu \nu}g_{\lambda \rho})
\right.$}
\Text(80,-6)[lb]{$ \left. +f_{adx}f_{xbc}
(g_{\mu \nu}g_{\lambda \rho}-g_{\mu \lambda}g_{\nu \rho}
-{\frac{1}{\xi}}g_{\mu \rho}g_{\nu \lambda})
\right.$}
\Text(80,-25)[lb]{$ \left. +f_{acx}f_{xbd}
(g_{\mu \nu}g_{\lambda \rho}-g_{\mu \rho}g_{\nu \lambda})
\right]$}
\end{picture}
\cput[]{00}(-1.01,0.65){\scalebox{0.5}{~~}}
\rput{00}(-0.99,0.67){A}
\cput[]{00}(0.16,-0.65){\scalebox{0.5}{~~}}
\rput{00}(0.15,-0.67){A}
\rput{00}(11.5,0.0){($\xi=\infty$ standard quartic vertex)}

\vspace*{9mm}

\begin{picture}(40,40)(0,0)
\SetWidth{1.2}
\Photon(19,0)(45,0){1}{4}
\DashLine(19,0)(19,+15){3}
\DashLine(19,0)(19,-15){3}
\Vertex(19,0){1.8}
\Text(19,+21)[]{a}
\Text(19,-21)[]{b}
\Text(13,+10)[]{p}
\Text(13,-10)[]{q}
\Text(46,-11)[]{c,$\mu$}
\Text(65,-6)[lb]{$-g f_{abc}\:(p-{q})_\mu $}
\end{picture}
\cput[]{00}(0.19,0){\scalebox{0.5}{~~}}
\rput{00}(0.18,0.02){A}
\rput{ 90}(-0.84,+.34){$>$}
\rput{-90}(-0.84,-.37){$<$}
\rput{00}(11.5,0.0){($q_\mu=0$ standard gluon--ghost vertex)}

\vspace*{9mm}

\begin{picture}(40,40)(0,0)
\SetWidth{1.2}
\Photon(30,0)(15,-15){1}{5}
\Photon(30,0)(45,-15){1}{5}
\DashLine(30,0)(15,15){3}
\DashLine(30,0)(45,15){3}
\Vertex(30,0){1.8}
\Text(15,21)[]{a}
\Text(15,-30)[]{c,$\mu$}
\Text(47,-20)[]{d,$\nu$}
\Text(45,21)[]{b}
\Text(65,-6)[lb]{$-ig^2 f_{acx}f_{xdb} g_{\mu \nu} $}
\end{picture}
\cput[]{00}(-1.03,-0.65){\scalebox{0.5}{~~}}
\rput{00}(-1.01,-0.67){A}
\rput{-45}(-0.82,+.37){$<$}
\rput{ 45}(-0.07,+.37){$<$}

\vspace*{1.2cm}

\begin{picture}(40,40)(0,0)
\SetWidth{1.2}
\Photon(30,0)(15,-15){1}{3}
\Photon(30,0)(45,-15){1}{3}
\DashLine(30,0)(15,15){3}
\DashLine(30,0)(45,15){3}
\Vertex(30,0){1.8}
\Text(15,21)[]{a}
\Text(15,-30)[]{c,$\mu$}
\Text(47,-30)[]{d,$\nu$}
\Text(45,21)[]{b}
\Text(65,-6)[lb]{$-ig^2 g_{\mu \nu}
(f_{acx}f_{xdb} +f_{adx}f_{xcb} )$}
\end{picture}
\cput[]{00}(-1.03,-0.65){\scalebox{0.5}{~~}}
\rput{00}(-1.01,-0.67){A}
\cput[]{00}(0.16,-0.65){\scalebox{0.5}{~~}}
\rput{00}(0.15,-0.67){A}
\rput{-45}(-0.82,+.37){$<$}
\rput{ 45}(-0.07,+.37){$<$}

\vspace{1.6cm}

All momenta are taken to be outgoing.

\newpage

{\Large \bf Appendix C: BF propagator diagrams.}\\[3mm]

a) Pure Yang--Mills contributions to the BF propagator~\cite{Abbott}\\[3mm]
\begin{center}
\begin{picture}(260,50)(-5,0)
\SetWidth{1.2}
\PhotonArc(50,0)(22,32,148){2}{7}
\Photon(10,25)(25,25){2}{3}
\Photon(75,25)(90,25){2}{3}
\DashCArc(50,25)(25,0,180){5}
\DashCArc(50,25)(25,180,360){5}
\Vertex(25,25){1.8}
\Vertex(75,25){1.8}
\Vertex(31,10){1.8}
\Vertex(69,10){1.8}
\PhotonArc(210,50)(22,212,328){2}{7}
\Photon(170,25)(185,25){2}{3}
\Photon(235,25)(250,25){2}{3}
\DashCArc(210,25)(25,0,180){5}
\DashCArc(210,25)(25,180,360){5}
\Vertex(185,25){1.8}
\Vertex(235,25){1.8}
\Vertex(191,40){1.8}
\Vertex(229,40){1.8}
\end{picture}

\vspace*{-14mm}

\rput{U}(-4.57,0.0){%
\cput[]{00}(0.4,0){\scalebox{0.5}{~~}}
\rput{00}(0.42,0.02){A}
\cput[]{00}(3.5,0){\scalebox{0.5}{~~}}
\rput{00}(3.52,0.02){A}
\rput{00}(2.0,+.9){$>$}
\rput{00}(2.0,-.86){$<$}
\cput[]{00}(6.0,0){\scalebox{0.5}{~~}}
\rput{00}(6.02,0.02){A}
\cput[]{00}(9.15,0){\scalebox{0.5}{~~}}
\rput{00}(9.17,0.02){A}
\rput{00}(7.6,+.9){$>$}
\rput{00}(7.6,-.86){$<$}
}

\vspace*{1.8cm}

\begin{picture}(260,50)(-5,0)
\SetWidth{1.2}
\PhotonArc(50,25)(25,180,360){2}{10}
\PhotonArc(50,25)(25,132,180){2}{3}
\PhotonArc(50,25)(25,0,48){2}{3}
\DashCArc(50,45)(15,0,180){5}
\DashCArc(50,45)(15,180,360){5}
\Photon(10,25)(25,25){2}{3}
\Photon(75,25)(90,25){2}{3}
\Vertex(25,25){1.8}
\Vertex(75,25){1.8}
\Vertex(35,45){1.8}
\Vertex(65,45){1.8}
\PhotonArc(210,25)(25,180,360){2}{10}
\PhotonArc(210,25)(25,132,180){2}{3}
\PhotonArc(210,25)(25,0,48){2}{3}
\PhotonArc(210,45)(15,0,180){2}{7}
\PhotonArc(210,45)(15,180,360){2}{7}
\Photon(170,25)(185,25){2}{3}
\Photon(235,25)(250,25){2}{3}
\Vertex(185,25){1.8}
\Vertex(235,25){1.8}
\Vertex(195,45){1.8}
\Vertex(225,45){1.8}
\end{picture}

\vspace*{-14mm}

\rput{U}(-4.57,0.0){%
\cput[]{00}(0.4,0){\scalebox{0.5}{~~}}
\rput{00}(0.42,0.02){A}
\cput[]{00}(3.5,0){\scalebox{0.5}{~~}}
\rput{00}(3.52,0.02){A}
\cput[]{00}(6.0,0){\scalebox{0.5}{~~}}
\rput{00}(6.02,0.02){A}
\cput[]{00}(9.15,0){\scalebox{0.5}{~~}}
\rput{00}(9.17,0.02){A}
\rput{00}(2.0,+1.24){$>$}
\rput{00}(2.0,+0.2){$<$}
}

\vspace*{1.8cm}

\begin{picture}(260,50)(-5,0)
\SetWidth{1.2}
\PhotonArc(50,25)(25,180,360){2}{10}
\DashCArc(50,25)(25,0,180){5}
\DashLine(25,25)(75,25){5}
\Photon(10,25)(25,25){2}{3}
\Photon(75,25)(90,25){2}{3}
\Vertex(25,25){1.8}
\Vertex(75,25){1.8}
\PhotonArc(210,25)(25,180,360){2}{10}
\PhotonArc(210,25)(25,000,180){2}{10}
\Photon(185,25)(235,25){2}{8}
\Photon(170,25)(185,25){2}{3}
\Photon(235,25)(250,25){2}{3}
\Vertex(185,25){1.8}
\Vertex(235,25){1.8}
\end{picture}

\vspace*{-14mm}

\rput{U}(-4.57,0.0){%
\cput[]{00}(0.4,0){\scalebox{0.5}{~~}}
\rput{00}(0.42,0.02){A}
\cput[]{00}(3.5,0){\scalebox{0.5}{~~}}
\rput{00}(3.52,0.02){A}
\cput[]{00}(6.0,0){\scalebox{0.5}{~~}}
\rput{00}(6.02,0.02){A}
\cput[]{00}(9.15,0){\scalebox{0.5}{~~}}
\rput{00}(9.17,0.02){A}
\rput{00}(2.0,+0.9){$>$}
\rput{00}(2.0,+0.0){$<$}
}

\vspace*{1.8cm}

\begin{picture}(260,50)(-5,0)
\SetWidth{1.2}
\PhotonArc(25,50)(25,270,360){2}{6}
\Photon(10,25)(25,25){2}{3}
\Photon(75,25)(90,25){2}{3}
\DashCArc(50,25)(25,0,180){5}
\DashCArc(50,25)(25,180,360){5}
\Vertex(25,25){1.8}
\Vertex(75,25){1.8}
\Vertex(50,50){1.8}
\PhotonArc(185,00)(22,000,090){2}{6}
\Photon(170,25)(185,25){2}{3}
\Photon(235,25)(250,25){2}{3}
\DashCArc(210,25)(25,0,180){5}
\DashCArc(210,25)(25,180,360){5}
\Vertex(185,25){1.8}
\Vertex(235,25){1.8}
\Vertex(210,00){1.8}
\end{picture}

\vspace*{-14mm}

\rput{U}(-4.57,0.0){%
\cput[]{00}(0.4,0){\scalebox{0.5}{~~}}
\rput{00}(0.42,0.02){A}
\cput[]{00}(3.5,0){\scalebox{0.5}{~~}}
\rput{00}(3.52,0.02){A}
\rput{-45}(2.0,0.0){%
\rput{00}(0.0,+.86){$>$}}
\rput{00}(2.0,-.9){$<$}
\cput[]{00}(6.0,0){\scalebox{0.5}{~~}}
\rput{00}(6.02,0.02){A}
\cput[]{00}(9.15,0){\scalebox{0.5}{~~}}
\rput{00}(9.17,0.02){A}
\rput{00}(7.6,+.9){$>$}
\rput{45}(7.6,0.0){%
\rput{00}(0.0,-.86){$<$}}
}

\vspace*{1.8cm}

\begin{picture}(260,50)(-5,0)
\SetWidth{1.2}
\PhotonArc(75,50)(25,180,270){2}{6}
\Photon(10,25)(25,25){2}{3}
\Photon(75,25)(90,25){2}{3}
\DashCArc(50,25)(25,0,180){5}
\DashCArc(50,25)(25,180,360){5}
\Vertex(25,25){1.8}
\Vertex(75,25){1.8}
\Vertex(50,50){1.8}
\PhotonArc(235,00)(22,090,180){2}{6}
\Photon(170,25)(185,25){2}{3}
\Photon(235,25)(250,25){2}{3}
\DashCArc(210,25)(25,0,180){5}
\DashCArc(210,25)(25,180,360){5}
\Vertex(185,25){1.8}
\Vertex(235,25){1.8}
\Vertex(210,00){1.8}
\end{picture}

\vspace*{-14mm}

\rput{U}(-4.57,0.0){%
\cput[]{00}(0.4,0){\scalebox{0.5}{~~}}
\rput{00}(0.42,0.02){A}
\cput[]{00}(3.5,0){\scalebox{0.5}{~~}}
\rput{00}(3.52,0.02){A}
\rput{45}(2.0,0.0){%
\rput{00}(0.0,+.90){$>$}}
\rput{00}(2.0,-.86){$<$}
\cput[]{00}(6.0,0){\scalebox{0.5}{~~}}
\rput{00}(6.02,0.02){A}
\cput[]{00}(9.15,0){\scalebox{0.5}{~~}}
\rput{00}(9.17,0.02){A}
\rput{00}(7.6,+.90){$>$}
\rput{-45}(7.6,0.0){%
\rput{00}(0.0,-.86){$<$}}
}

\vspace*{1.8cm}

\begin{picture}(260,50)(-5,0)
\SetWidth{1.2}
\PhotonArc(50,25)(25,180,360){2}{10}
\PhotonArc(50,25)(25,0,130){2}{7}
\DashCArc(31.0,35.4)(12,0,180){4}
\DashCArc(31.0,35.4)(12,180,360){4}
\Photon(10,25)(25,25){2}{3}
\Photon(75,25)(90,25){2}{3}
\Vertex(25,25){1.8}
\Vertex(75,25){1.8}
\Vertex(35,45){1.8}
\PhotonArc(210,25)(25,0,130){2}{7}
\PhotonArc(210,25)(25,180,360){2}{10}
\PhotonArc(191,35.4)(12,0,180){2}{6}
\PhotonArc(191,35.4)(12,180,360){2}{6}
\Photon(170,25)(185,25){2}{3}
\Photon(235,25)(250,25){2}{3}
\Vertex(185,25){1.8}
\Vertex(235,25){1.8}
\Vertex(195,45){1.8}
\end{picture}

\vspace*{-14mm}

\rput{U}(-4.57,0.0){%
\cput[]{00}(0.4,0){\scalebox{0.5}{~~}}
\rput{00}(0.42,0.02){A}
\cput[]{00}(3.5,0){\scalebox{0.5}{~~}}
\rput{00}(3.52,0.02){A}
\cput[]{00}(6.0,0){\scalebox{0.5}{~~}}
\rput{00}(6.02,0.02){A}
\cput[]{00}(9.15,0){\scalebox{0.5}{~~}}
\rput{00}(9.17,0.02){A}
\rput{00}(0.4,0.0){%
\rput{60}(0.5,+0.5){$>$}
\rput{60}(1.2,+0.1){$<$}
}}

\vspace*{1.8cm}

\begin{picture}(260,50)(-5,0)
\SetWidth{1.2}
\PhotonArc(50,25)(25,180,360){2}{10}
\PhotonArc(50,25)(25,50,180){2}{7}
\DashCArc(69.0,35.4)(12,0,180){4}
\DashCArc(69.0,35.4)(12,180,360){4}
\Photon(10,25)(25,25){2}{3}
\Photon(75,25)(90,25){2}{3}
\Vertex(25,25){1.8}
\Vertex(75,25){1.8}
\Vertex(66,46){1.8}
\PhotonArc(210,25)(25,50,180){2}{7}
\PhotonArc(210,25)(25,180,360){2}{10}
\PhotonArc(229,35.4)(12,0,180){2}{6}
\PhotonArc(229,35.4)(12,180,360){2}{6}
\Photon(170,25)(185,25){2}{3}
\Photon(235,25)(250,25){2}{3}
\Vertex(185,25){1.8}
\Vertex(235,25){1.8}
\Vertex(223,45){1.8}
\end{picture}

\vspace*{-14mm}

\rput{U}(-4.57,0.0){%
\cput[]{00}(0.4,0){\scalebox{0.5}{~~}}
\rput{00}(0.42,0.02){A}
\cput[]{00}(3.5,0){\scalebox{0.5}{~~}}
\rput{00}(3.52,0.02){A}
\cput[]{00}(6.0,0){\scalebox{0.5}{~~}}
\rput{00}(6.02,0.02){A}
\cput[]{00}(9.15,0){\scalebox{0.5}{~~}}
\rput{00}(9.17,0.02){A}
\rput{-45}(2.6,0.4){%
\rput{00}(0.0,+.42){$>$}
\rput{00}(0.0,-.43){$<$}}
}

\vspace*{1.8cm}

\begin{picture}(260,50)(-5,0)
\SetWidth{1.2}
\PhotonArc(50,25)(25,180,360){2}{10}
\PhotonArc(50,25)(25,000,180){2}{10}
\Photon(50,00)(50,50){2}{8}
\Photon(10,25)(25,25){2}{3}
\Photon(75,25)(90,25){2}{3}
\Vertex(25,25){1.8}
\Vertex(75,25){1.8}
\Vertex(50,50){1.8}
\Vertex(50,00){1.8}
\Photon(210,00)(210,50){2}{8}
\DashCArc(210,25)(25,0,180){5}
\DashCArc(210,25)(25,180,360){5}
\Photon(170,25)(185,25){2}{3}
\Photon(235,25)(250,25){2}{3}
\Vertex(185,25){1.8}
\Vertex(235,25){1.8}
\Vertex(210,50){1.8}
\Vertex(210,00){1.8}
\end{picture}

\vspace*{-14mm}

\rput{U}(-4.57,0.0){%
\cput[]{00}(0.4,0){\scalebox{0.5}{~~}}
\rput{00}(0.42,0.02){A}
\cput[]{00}(3.5,0){\scalebox{0.5}{~~}}
\rput{00}(3.52,0.02){A}
\cput[]{00}(6.0,0){\scalebox{0.5}{~~}}
\rput{00}(6.02,0.02){A}
\cput[]{00}(9.15,0){\scalebox{0.5}{~~}}
\rput{00}(9.17,0.02){A}
\rput{45}(7.6,0.0){%
\rput{00}(0.0,+.9){$>$}}
\rput{45}(7.6,0.0){%
\rput{00}(0.0,-.86){$<$}}
}
\end{center}

\newpage

\begin{center}
\begin{picture}(260,50)(-5,0)
\SetWidth{1.2}
\PhotonArc(50,25)(25,90,270){2}{10}
\DashCArc(50,25)(25,270,090){5}
\DashLine(50,00)(50,50){5}
\Photon(10,25)(25,25){2}{3}
\Photon(75,25)(90,25){2}{3}
\Vertex(25,25){1.8}
\Vertex(75,25){1.8}
\Vertex(50,50){1.8}
\Vertex(50,00){1.8}
\PhotonArc(210,25)(25,270,090){2}{10}
\DashCArc(210,25)(25,090,270){5}
\DashLine(210,00)(210,50){5}
\Photon(170,25)(185,25){2}{3}
\Photon(235,25)(250,25){2}{3}
\Vertex(185,25){1.8}
\Vertex(235,25){1.8}
\Vertex(210,50){1.8}
\Vertex(210,00){1.8}
\end{picture}

\vspace*{-14mm}

\rput{U}(-4.57,0.0){%
\cput[]{00}(0.4,0){\scalebox{0.5}{~~}}
\rput{00}(0.42,0.02){A}
\cput[]{00}(3.5,0){\scalebox{0.5}{~~}}
\rput{00}(3.52,0.02){A}
\cput[]{00}(6.0,0){\scalebox{0.5}{~~}}
\rput{00}(6.02,0.02){A}
\cput[]{00}(9.15,0){\scalebox{0.5}{~~}}
\rput{00}(9.17,0.02){A}
\rput{45}(7.54,0.0){%
\rput{00}(0.0,+.86){$>$}}
\rput{-45}(7.54,0.0){%
\rput{00}(0.0,-.86){$<$}}
\rput{-45}(2.0,0.0){%
\rput{00}(0.0,+.86){$>$}}
\rput{45}(2.0,0.0){%
\rput{00}(0.0,-.86){$<$}}
\rput{00}(1.94,0.0){\rput{-90}(0,0){$<$}}
\rput{00}(7.6,0.0){\rput{90}(0,0){$<$}}
}

\vspace*{1.8cm}

\begin{picture}(260,50)(-5,0)
\SetWidth{1.2}
\PhotonArc(37.5,25)(12.5,180,360){2}{6}
\PhotonArc(37.5,25)(12.5,000,180){2}{6}
\PhotonArc(62.5,25)(12.5,180,360){2}{6}
\PhotonArc(62.5,25)(12.5,000,180){2}{6}
\Photon(10,25)(25,25){2}{3}
\Photon(75,25)(90,25){2}{3}
\Vertex(25,25){1.8}
\Vertex(75,25){1.8}
\Vertex(50,25){1.8}
\PhotonArc(210,25)(25,180,360){2}{10}
\PhotonArc(210,25)(25,000,180){2}{10}
\Photon(170,25)(185,25){2}{3}
\Photon(235,25)(250,25){2}{3}
\Vertex(185,25){1.8}
\Vertex(235,25){1.8}
\Vertex(210,50){1.8}
\GText(210,50){0.5}{·}
\end{picture}

\vspace*{-14mm}

\rput{U}(-4.57,0.0){%
\cput[]{00}(0.4,0){\scalebox{0.5}{~~}}
\rput{00}(0.42,0.02){A}
\cput[]{00}(3.5,0){\scalebox{0.5}{~~}}
\rput{00}(3.52,0.02){A}
\cput[]{00}(6.0,0){\scalebox{0.5}{~~}}
\rput{00}(6.02,0.02){A}
\cput[]{00}(9.15,0){\scalebox{0.5}{~~}}
\rput{00}(9.17,0.02){A}
}

\vspace*{1.8cm}

\begin{picture}(260,50)(-5,0)
\SetWidth{1.2}
\PhotonArc(50,25)(25,180,360){2}{10}
\PhotonArc(50,25)(25,000,180){2}{10}
\Photon(10,25)(25,25){2}{3}
\Photon(75,25)(90,25){2}{3}
\Vertex(25,25){1.8}
\Vertex(75,25){1.8}
\GText(25,25){0.5}{·}
\PhotonArc(210,25)(25,180,360){2}{10}
\PhotonArc(210,25)(25,000,180){2}{10}
\Photon(170,25)(185,25){2}{3}
\Photon(235,25)(250,25){2}{3}
\Vertex(185,25){1.8}
\Vertex(235,25){1.8}
\GText(235,25){0.5}{·}
\end{picture}

\vspace*{-14mm}

\rput{U}(-4.57,0.0){%
\cput[]{00}(0.4,0){\scalebox{0.5}{~~}}
\rput{00}(0.42,0.02){A}
\cput[]{00}(3.5,0){\scalebox{0.5}{~~}}
\rput{00}(3.52,0.02){A}
\cput[]{00}(6.0,0){\scalebox{0.5}{~~}}
\rput{00}(6.02,0.02){A}
\cput[]{00}(9.15,0){\scalebox{0.5}{~~}}
\rput{00}(9.17,0.02){A}
}
\end{center}

\vspace*{1.8cm}

b) Fermionic contributions to BF propagator~\cite{AGS}\\[3mm]

\begin{center}
\begin{picture}(260,50)(-5,0)
\SetWidth{1.2}
\PhotonArc(50,25)(25,180,360){2}{10}
\PhotonArc(50,25)(25,132,180){2}{3}
\PhotonArc(50,25)(25,0,48){2}{3}
\CArc(50,45)(15,0,180)
\CArc(50,45)(15,180,360)
\Photon(10,0)(90,0){2}{10}
\Vertex(50,0){3.2}
\Vertex(35,45){1.8}
\Vertex(65,45){1.8}
\PhotonArc(210,25)(25,180,360){2}{10}
\PhotonArc(210,25)(25,132,180){2}{3}
\PhotonArc(210,25)(25,0,48){2}{3}
\CArc(210,45)(15,0,180)
\CArc(210,45)(15,180,360)
\Photon(170,25)(185,25){2}{3}
\Photon(235,25)(250,25){2}{3}
\Vertex(185,25){1.8}
\Vertex(235,25){1.8}
\Vertex(195,45){1.8}
\Vertex(225,45){1.8}
\end{picture}

\vspace*{-14mm}

\rput{U}(-4.57,0.0){%
\cput[]{00}(0.4,-.86){\scalebox{0.5}{~~}}
\rput{00}(0.42,-.84){A}
\cput[]{00}(3.5,-.86){\scalebox{0.5}{~~}}
\rput{00}(3.52,-.84){A}
\cput[]{00}(6.0,0){\scalebox{0.5}{~~}}
\rput{00}(6.02,0.02){A}
\cput[]{00}(9.15,0){\scalebox{0.5}{~~}}
\rput{00}(9.17,0.02){A}
\rput{00}(2.0,+1.24){$>$}
\rput{00}(2.0,+0.2){$<$}
\rput{00}(7.6,+1.24){$>$}
\rput{00}(7.6,+0.2){$<$}
}

\vspace*{1.8cm}

\begin{picture}(260,50)(-5,0)
\SetWidth{1.2}
\PhotonArc(50,0)(22,32,148){2}{7}
\Photon(10,25)(25,25){2}{3}
\Photon(75,25)(90,25){2}{3}
\CArc(50,25)(25,0,180)
\CArc(50,25)(25,180,360)
\Vertex(25,25){1.8}
\Vertex(75,25){1.8}
\Vertex(31,10){1.8}
\Vertex(69,10){1.8}
\PhotonArc(210,50)(22,212,328){2}{7}
\Photon(170,25)(185,25){2}{3}
\Photon(235,25)(250,25){2}{3}
\CArc(210,25)(25,0,180)
\CArc(210,25)(25,180,360)
\Vertex(185,25){1.8}
\Vertex(235,25){1.8}
\Vertex(191,40){1.8}
\Vertex(229,40){1.8}
\end{picture}

\vspace*{-14mm}

\rput{U}(-4.57,0.0){%
\cput[]{00}(0.4,0){\scalebox{0.5}{~~}}
\rput{00}(0.42,0.02){A}
\cput[]{00}(3.5,0){\scalebox{0.5}{~~}}
\rput{00}(3.52,0.02){A}
\rput{00}(2.0,+.9){$>$}
\rput{00}(2.0,-.86){$<$}
\cput[]{00}(6.0,0){\scalebox{0.5}{~~}}
\rput{00}(6.02,0.02){A}
\cput[]{00}(9.15,0){\scalebox{0.5}{~~}}
\rput{00}(9.17,0.02){A}
\rput{00}(7.6,+.9){$>$}
\rput{00}(7.6,-.86){$<$}
}

\vspace*{1.8cm}

\begin{picture}(260,50)(-5,0)
\SetWidth{1.2}
\PhotonArc(50,25)(25,90,270){2}{10}
\CArc(50,25)(25,270,090)
\Line(50,00)(50,50)
\Photon(10,25)(25,25){2}{3}
\Photon(75,25)(90,25){2}{3}
\Vertex(25,25){1.8}
\Vertex(75,25){1.8}
\Vertex(50,50){1.8}
\Vertex(50,00){1.8}
\PhotonArc(210,25)(25,270,090){2}{10}
\CArc(210,25)(25,090,270)
\Line(210,00)(210,50)
\Photon(170,25)(185,25){2}{3}
\Photon(235,25)(250,25){2}{3}
\Vertex(185,25){1.8}
\Vertex(235,25){1.8}
\Vertex(210,50){1.8}
\Vertex(210,00){1.8}
\end{picture}

\vspace*{-14mm}

\rput{U}(-4.57,0.0){%
\cput[]{00}(0.4,0){\scalebox{0.5}{~~}}
\rput{00}(0.42,0.02){A}
\cput[]{00}(3.5,0){\scalebox{0.5}{~~}}
\rput{00}(3.52,0.02){A}
\cput[]{00}(6.0,0){\scalebox{0.5}{~~}}
\rput{00}(6.02,0.02){A}
\cput[]{00}(9.15,0){\scalebox{0.5}{~~}}
\rput{00}(9.17,0.02){A}
\rput{45}(7.54,0.0){%
\rput{00}(0.0,+.86){$>$}}
\rput{-45}(7.54,0.0){%
\rput{00}(0.0,-.86){$<$}}
\rput{-45}(2.0,0.0){%
\rput{00}(0.0,+.86){$>$}}
\rput{45}(2.0,0.0){%
\rput{00}(0.0,-.86){$<$}}
\rput{00}(1.94,0.0){\rput{-90}(0,0){$<$}}
\rput{00}(7.6,0.0){\rput{90}(0,0){$<$}}
}

\vspace*{1.8cm}

\begin{picture}(260,50)(-5,0)
\SetWidth{1.2}
\Photon(50,00)(50,50){2}{8}
\CArc(50,25)(25,180,360)
\CArc(50,25)(25,000,180)
\Photon(10,25)(25,25){2}{3}
\Photon(75,25)(90,25){2}{3}
\Vertex(25,25){1.8}
\Vertex(75,25){1.8}
\Vertex(50,50){1.8}
\Vertex(50,00){1.8}
\PhotonArc(210,25)(25,180,360){2}{10}
\PhotonArc(210,25)(25,000,180){2}{10}
\Photon(170,25)(185,25){2}{3}
\Photon(235,25)(250,25){2}{3}
\Vertex(185,25){1.8}
\Vertex(235,25){1.8}
\Vertex(210,50){1.8}
\Text(210,50)[]{\Huge \liste\symbol{'156}}
\end{picture}

\vspace*{-14mm}

\rput{U}(-4.57,0.0){%
\cput[]{00}(0.4,0){\scalebox{0.5}{~~}}
\rput{00}(0.42,0.02){A}
\cput[]{00}(3.5,0){\scalebox{0.5}{~~}}
\rput{00}(3.52,0.02){A}
\cput[]{00}(6.0,0){\scalebox{0.5}{~~}}
\rput{00}(6.02,0.02){A}
\cput[]{00}(9.15,0){\scalebox{0.5}{~~}}
\rput{00}(9.17,0.02){A}
\rput{45}(2.0,0.0){%
\rput{00}(0.0,+.9){$>$}}
\rput{45}(2.0,0.0){%
\rput{00}(0.0,-.86){$<$}}
}

\vspace*{1.8cm}

\begin{picture}(260,50)(-5,0)
\SetWidth{1.2}
\PhotonArc(50,25)(25,180,360){2}{10}
\PhotonArc(50,25)(25,000,180){2}{10}
\Photon(10,25)(25,25){2}{3}
\Photon(75,25)(90,25){2}{3}
\Vertex(25,25){1.8}
\Vertex(75,25){1.8}
\Text(25,25)[]{\Huge \liste\symbol{'156}}
\PhotonArc(210,25)(25,180,360){2}{10}
\PhotonArc(210,25)(25,000,180){2}{10}
\Photon(170,25)(185,25){2}{3}
\Photon(235,25)(250,25){2}{3}
\Vertex(185,25){1.8}
\Vertex(235,25){1.8}
\Text(235,25)[]{\Huge \liste\symbol{'156}}
\end{picture}

\vspace*{-14mm}

\rput{U}(-4.57,0.0){%
\cput[]{00}(0.4,0){\scalebox{0.5}{~~}}
\rput{00}(0.42,0.02){A}
\cput[]{00}(3.5,0){\scalebox{0.5}{~~}}
\rput{00}(3.52,0.02){A}
\cput[]{00}(6.0,0){\scalebox{0.5}{~~}}
\rput{00}(6.02,0.02){A}
\cput[]{00}(9.15,0){\scalebox{0.5}{~~}}
\rput{00}(9.17,0.02){A}
}
\end{center}

\end{document}